\documentclass[english,a4paper,prx,twocolumn,amsmath,amssymb,superscriptaddress]{revtex4}
\usepackage{graphicx,color,natbib,subfigure}
\usepackage{amsmath,amssymb}
\usepackage[dvipsnames]{xcolor}

\newcommand{\pecl}{\operatorname{\mathit{P\kern-.08em e}}}

\begin{document}

\title{Coupling of sedimentation and liquid structure: influence on hard sphere nucleation}

\author{Nicholas Wood}
\affiliation{HH Wills Physics Laboratory, Tyndall Avenue, Bristol, BS8 1TL, UK}
\affiliation{Centre for Nanoscience and Quantum Information, Tyndall Avenue, Bristol, BS8 1FD, UK}
\author{John Russo}
\affiliation{School of Mathematics, University Walk, Bristol, BS8 1TW, UK}
\author{Francesco Turci}
\affiliation{HH Wills Physics Laboratory, Tyndall Avenue, Bristol, BS8 1TL, UK}
\affiliation{Centre for Nanoscience and Quantum Information, Tyndall Avenue, Bristol, BS8 1FD, UK}
\author{C. Patrick Royall}
\affiliation{HH Wills Physics Laboratory, Tyndall Avenue, Bristol, BS8 1TL, UK}
\affiliation{Centre for Nanoscience and Quantum Information, Tyndall Avenue, Bristol, BS8 1FD, UK}
\affiliation{School of Chemistry, University of Bristol, Cantock Close, Bristol, BS8 1TS, UK}

\begin{abstract}
The discrepancy in nucleation rate densities between simulated and experimental hard spheres remains staggering and unexplained. Suggestively, more strongly sedimenting colloidal suspensions of hard spheres nucleate much faster than weakly sedimenting systems. In this work we consider firstly the effect of sedimentation on the structure of colloidal hard spheres, by tuning the density mismatch between solvent and colloidal particles. In particular we investigate the effect on the degree of five fold symmetry present. Secondly we study the size of density fluctuations in these experimental systems in comparison to simulations. The density fluctuations are measured by assigning each particle a local density, which is related to the number of particles within a distance of 3.25 particle diameters. The standard deviation of these local densities gives an indication of the fluctuations present in the system. Five fold symmetry is suppressed by a factor of two when sedimentation is induced in our system. Density fluctuations are also increased by a factor of two in experiments compared to simulations. The change in five fold symmetry makes a difference to the expected nucleation rates, but we demonstrate that it is ultimately too small to resolve the discrepancy between experiment and simulation, while the fluctuations are shown to be an artefact of 3d particle tracking.
\end{abstract}


\maketitle

\section{Introduction}

Nucleation is a long-standing challenge, due not least to its local nature, where rare events on microscopic time and length-scales initiate the macroscopic phase transition \cite{sear2007}. Bridging these time and length-scales is a formidable task, and one system has emerged as an ideal testing ground for a comparison of theory and experiment. This is the hard sphere model, prized for its simplicity, yet capturing the essential thermodynamics of nucleation. Systems very close to hard spheres can be realised with colloidal suspensions\cite{schilling2011,hunter2012,royall2013myth,palberg2014,gasser2001,gasser2009}, which can be imaged at the particle level, providing an unprecedented opportunity to compare prediction and reality \cite{ivlev}.

Predicting rare events is the key to a wide range of important problems from rate determination in chemical reactions to absorption of drug molecules by cells \cite{allen2009}, in addition to nucleation. Advanced rare event sampling techniques have been developed which have been applied to a huge variety of these types of problems \cite{ghaemi2012,bussi2006}. We would expect to find an ideal experimental test case for these techniques in the nucleation rates of colloidal suspensions of hard spheres, due to the simplicity of the model. Alas, the landmark work which introduced rare event sampling to hard sphere nucleation revealed a catastrophic divergence from the experimental results, with discrepancies \emph{up to 20 orders of magnitude in the nucleation rate} \cite{auer2001}. Over a decade on, despite numerous attempts to address possible causes \cite{auer2004,filion2011,palberg2014}, the discrepancy persists. Because rare event sampling is used so widely throughout science \cite{ghaemi2012,bussi2006}, and because the hard sphere system represents the best case scenario for testing theoretical and computational predictions, the importance of this issue reaches far beyond hard spheres and indeed far beyond crystallisation in general. 

In short, either the experimental measurements or the numerical predictions must exhibit some profound flaw, or \emph{they do not pertain to the same system}. Previous experimental work which considered the size polydispersity of the colloidal particles \cite{auer2001}, electrostatic charge \cite{auer2002}, hydrodynamic coupling between the colloids due to the solvent \cite{radu2014} and sedimentation \emph{in the absence of hydrodynamics} \cite{russo2013} have not resolved the disagreement. Meanwhile, whilst the simulation results have been reproduced with different rare event sampling techniques \cite{filion2011} questions linger over the use of classical nucleation theory and whether the correct reaction co-ordinate has been chosen \cite{sear2012}.

\begin{figure}
\centering
\includegraphics[width=90mm]{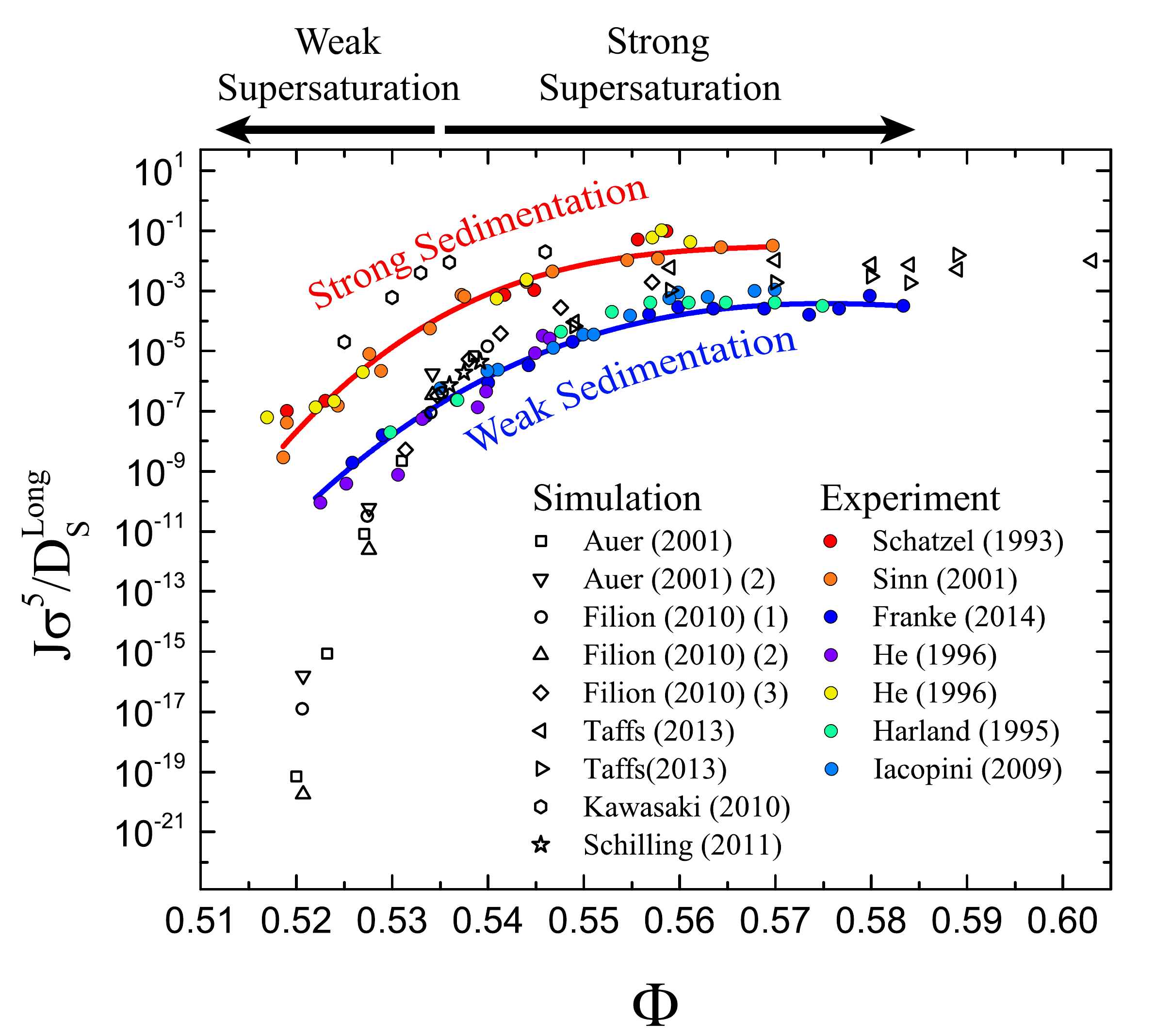}
\caption{\textbf{Literature Nucleation Rates.} A comparison of literature hard sphere nucleation rates as a function of volume fraction $\phi$. Scaling by $\sigma^5$/$D_0$ gives a dimensionless quantity which can be used to compare different experimental systems and simulation. Coloured circles are experimental data, whilst unfilled shapes are simulation data. The experimental data fall into two regimes, one group of experiments with Pecl\'{e}t numbers of around 0.2 nucleate very quickly, whilst the second group of experiments agree more closely with simulations and have a Pecl\'{e}t number of around 0.01. This figure is based on references\cite{schilling2011,palberg2014}.
}
\label{figLiteratureNucleationRate}
\end{figure}

The discrepancy emerges at weak supersaturation --- when nucleation is a rare event. At increased supersaturation, when the nucleation rate is much faster, little discrepancy is found \cite{schilling2011,palberg2014,taffs2013}. Careful examination of the experimental literature \cite{schilling2011,palberg2014} reveals two trends in the experimental data as indicated in Fig.~\ref{figLiteratureNucleationRate}. Work with colloids which sediment relatively rapidly falls on roughly same curve of nucleation as a function of supersaturation \cite{taffs2013,schatzel1993,he1996,sinn2001} while other experiments, which sediment relatively slowly, exhibit very reduced rates of sedimentation and appear to be somewhat closer to the numerical predictions \cite{harland1997,iacopini2009,frankethesis,he1996,taffs2013}. Both sets of data extend to regime of weak supersaturation in which nucleation is a rare event and the discrepancy is found. Here we characterise sedimentation by the gravitational Pecl\'{e}t number $Pe=\tau_B/\tau_\mathrm{sed}$ which compares the time taken by a particle to diffuse its own radius $\sigma$, $\tau_B=\sigma^3 \pi \eta/8 k_BT$ (where $\eta$ is the viscosity of immersing solvent and $k_BT$ is the thermal energy) and the time to sediment its radius $\tau_\mathrm{sed}=3 \pi \eta \sigma^2/2\delta m g$ where $\delta m$ is Archimedes' bouyant mass and $g$ is the acceleration due to gravity.

In 1952 Sir Charles Frank postulated that the presence of fivefold-symmetric icosahedra in supercooled fluids suppressed crystallisation, allowing the formation of the glass \cite{frank1952}. We have previously shown that the relationship between crystallisation and five fold symmetry is directly causative, showing that the suppression of five fold symmetry increases the nucleation rate by several orders of magnitude by decreasing the fluid-crystal surface tension \cite{taffs2016}. The proxy for five fold symmetry we choose to study here is the \emph{defective icosahedron} which is equivalent to the icosahedron with 3 outer particles removed \cite{malins2013tcc}. This structure retains 3 of the five fold symmetric rings contained in the full 13-membered icosahedron (see Fig.~\ref{figVisualisation}). The defective icosahedron is a common motif in supercooled hard sphere fluids \cite{royall2015,royall2017,royall2018} and has been related to their slow dynamics \cite{royall2015,royall2017}. Moreover, its population decreases during the crystallisation process and this is concomitant with the rise of the FCC crystal \cite{taffs2013,taffs2016}.

Here we take a two-pronged approach to probe which aspect of sedimentation may be responsible for the discrepancy. Firstly, we use carefully controlled experiments on nearly hard sphere colloids to demonstrate the effect of sedimentation on the structure of metastable fluids at weak supersaturation in the regime in which the discrepancy is found. Our results reveal a significant reduction in fivefold symmetry in the case that the system undergoes sedimentation. Then we investigate this effect of sedimentation on the barrier height to nucleation using umbrella sampling for a system in which the degree of fivefold symmetry is controlled. For this purpose we employ umbrella sampling of systems with varying degrees of fivefold symmetry. Our results reveal the potential for a very considerable change in nucleation rates, although this falls somewhat short of explaining the entire discrepancy. Secondly we examine the extent to which the local density fluctuations of the experiments differ from those in the simulations. We find a large difference here as well, with many particles having much higher surrounding density in the experiments than in the simulations. This, however, we show to be an artefact of particle tracking, rather than an effect present in the real colloidal fluid.

We carry out particle-resolved studies of a well-controlled model system of polymethyl methacrylate colloids of diameter 2$\mu$m, which are fluorescently labelled. We suspend the particles in a solvent mixture of cis-decalin and cyclohexyl bromide (CHB), in which 4 mMol of tetrabutyl ammonium bromide salt is dissolved to suppress the electrostatic interactions. We have previously shown that this system closely resembles the hard sphere model \cite{taffs2013}. Here we control sedimentation by varying the degree of density matching through varying the solvent composition, making it either rich in CHB (colloids move upwards) or in cis-decalin (colloids sediment). To determine the degree of fivefold symmetry in our system, we use the topological cluster classification (TCC) ~\cite{malins2013tcc}, which directly identifies ten-membered defective icosahedra.

\section{Methods}
\subsection{Experimental}\label{expmethods}

We employ PMMA spheres of diameter 2.0 microns and a polydispersity of around 4\% as determined by scanning electron microscopy. These spheres are suspended in a mixture of cyclohexyl bromide (CHB) and cis-decalin. This system allows the refractive index to be matched, whilst also giving control over the Pecl\'{e}t number through finely tuning the density mismatch between the solvent mixture and the particles. We focused on systems that are both well density matched and poorly density matched. The standard criterion employed for good density matching is that the suspension should show no sign of sedimentation after 30 minutes of centrifugation at 13400 RPM.

Care must be taken when changing the solvent composition to study the effect of sedimentation. Changing the solvent composition also changes the dielectric constant, and with it the electrostatics of the system; this alteration of the interparticle interactions may affect the structure of the fluid, and hence the nucleation rate. We therefore performed two density mismatched experiments, the first with particles moving against gravity with the addition of CHB to the density matched solvent (creaming), the second with particles moving in the direction of gravity (sedimenting), with the addition of cis-decalin. We know that the hydrodynamics of sedimentation is symmetric - the same hydrodynamics apply regardless of the direction of the bulk motion of the colloids through the solvent. However, the electrostatic effects are not symmetric about this change. In the case of CHB rich samples, we expect electrostatic charging to be stronger due to the higher dielectric constant. This effect is in favour of larger populations of higher-order structure. In the case of the cis-decalin samples, the electrostatic charging will be weaker  which will suppress populations of higher-order structure.
This experimental design therefore provides strong evidence for the cause of a given change in local structures between the density matched case and the sedimenting case --- a hydrodynamic effect will be unaffected by reversing the direction of motion, an electrostatic or imaging effect will not.

The poor density matching cases had solvents of composition 99\% CHB, 1\% cis decalin (Pe $\approx$ 1.5) (creaming), and 66\% cis decalin and 33\% CHB (Pe $\approx$ 1.5) (sedimenting). These samples are imaged in 0.5mm square capillaries using a Leica SP8. Particle tracking is performed in the bulk far from the capillary walls to avoid the influence of any potential heterogeneously nucleating crystals. Volume fractions are calculated by taking the ratio of the volume of identified particles to the size of the imaging box. The obtained volume fractions are consistent with those estimated through the fitting of the radial distribution function with the theoretical predictions of the Percus-Yevick theory for hard spheres \cite{hansen}.

When analysing bond networks with the TCC, there are choices to be made about what accurately constitutes a bond. One parameter involved here is the so-called $f_c$ parameter, which controls the condition our modified Voronoi analysis applies when identifying four membered rings. For the experimental data, we set this parameter to 0.87, in line with previous experimental work \cite{pinchaipat2017,hallett2018}, whilst for the simulation data it is set to 0.82. This relaxation for the experimental data takes into account the fact that in experiments there is an error on each identified particle centre position, whilst in simulation all particle positions are known perfectly \cite{royall2013myth}. These errors tend to reduce populations of defective icosahedra, especially the larger structures we are interested in, by artificially breaking Voronoi bonds. This correction only changes the total populations of defective icosahedra, any differences between populations in density matched samples and populations in sedimenting samples is robust with respect to the choice of the $f_c$ parameter.

\subsection{Simulations}

We employ Monte Carlo simulations in the isothermal-isochoric (NVT) and isothermal-isobaric (NPT) ensembles. Data is presented for the NVT ensemble unless otherwise stated. To approximate Brownian dynamics, we limit the moves to small, single particle displacements with a maximum step size of $0.04\,\sigma$ \cite{sanz2010}. Our hard sphere system is biased towards or away from the formation of pentagonal bipyramids using the scheme developed in Ref. 27. Here structures are identified via common neighbour analysis\cite{honeycutt1987} .
The pentagonal bipyramid ``1551'' structure comprises a bonded ``spindle'' pair of particles which share five, and only five, neighbours. There are five bonds in total between these five neighbours, which form a five-membered ring. We identify bonds using a Voronoi method with a maximum bond length set to $1.4\,\sigma$, a value chosen to capture the relevant bond network \cite{malins2013tcc}. Biasing is achieved by associating the formation of each pentagonal bipyramid with an energy penalty or reward of $\varepsilon\,k_{B}T$. That is, the field strength $\varepsilon$ denotes the energy change associated with the formation of each individual pentagonal bipyramid structure. 
A negative $\varepsilon$ encourages pentagonal bipyramid formation, a positive $\varepsilon$ suppresses pentagonal bipyramid formation. Each new trial configuration is then accepted or rejected according to the Metropolis algorithm.

Nucleation barriers are computed with a variant of the Umbrella Sampling technique, called CNT-US~\cite{rovigatti2018}, allowing the simulations to sample rare fluctuations comprising nuclei of critical size. This equilibrium sampling allows us to determine the crystal size distribution $N(n)$ (where $n$ is the nucleus size), and from this obtain the free energy barrier as $\beta\Delta G(n)=-k_BT\log N(n)$.

\section{Results}

Our analysis is divided into two sections. Firstly we examine the effect of sedimentation on the degree of five-fold symmetry in our colloidal systems, and the consequences for the nucleation rates. We then examine the effect of sedimentation on the size of the density fluctuations within the fluid.

Before discussing our experimental results, we emphasise the need for care in interpreting experimental data. As indicated in Fig.~\ref{figLiteratureNucleationRate}, the nucleation rate can vary by many orders of magnitude, for a tiny change in the control parameter, volume fraction. Worse, determining the volume fraction accurately is extremely hard, with typical errors in the absolute volume fraction of $\sim 0.03$ for these conditions \cite{poon2012}, which may be further compounded by the effects of electrostatic charging and other deviations of ideal hard sphere behaviour \cite{royall2013myth}. Therefore, here we have carefully calibrated our experiments by reference to previous experimental and simulation data. Specifically, we used the same solvent system as that used previously where the electrostatics in particular were carefully monitored \cite{taffs2013}. We checked that the crystallisation times were the same as those obtained previously at the same volume fraction. Crystallisation times were measured as the time for a large section of the sample to crystallise, determined by direct inspection of the confocal images. These experiments were performed in capillaries with a highly polydisperse PMMA of diameter 3.5 microns sintered to the walls to suppress heterogeneous nucleation. Comparing a non-equilibrium quantity such as the crystallisation time gives us a high degree of confidence that our experimental volume fractions are in agreement with previous work.

\subsection{Sedimentation and Five Fold Symmetry} \label{10B}

\begin{figure}
\centering
\includegraphics[width=75mm]{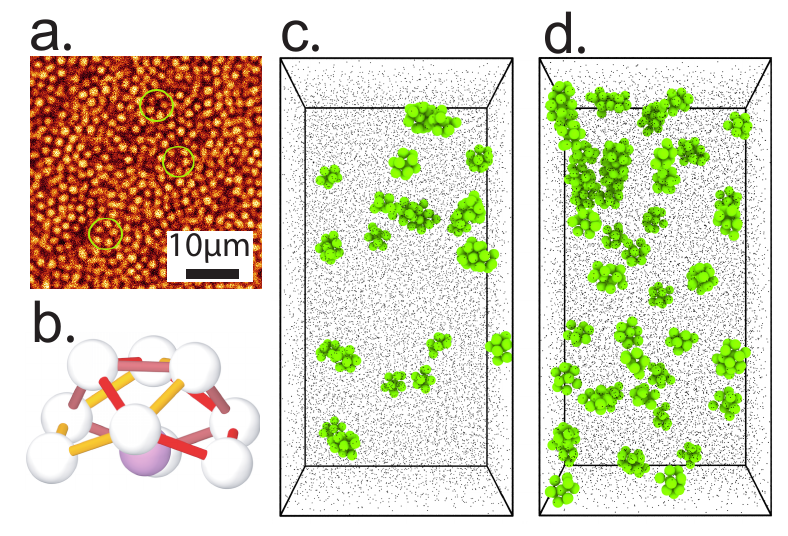}
\caption{\textbf{a.} A single slice through a 3d confocal image stack. 3 five membered rings are circled. Each of our full confocal stacks is comprised of 500 images like this, each taken at a different height in the capillary. These images allow us to build up a full 3 dimensional image of the sample. \textbf{b.} Diagram of the defective icosahedron (10B). Five membered rings are indicated in yellow, red, and dark red. \textbf{c,d.} Visualisation of the effect of sedimentation upon the structure of the metastable hard sphere fluid. \textbf{c} shows a sedimenting system, \textbf{d} shows a density matched system.  
This is \emph{experimental} data 
rendered 
after all particle centres have been located with particle tracking. Green particles are those found in defective icosahedra, particles not in defective icosahedra are rendered as smaller grey points for clarity. These snapshots are at a volume fraction of $\phi=0.45$. 
}
\label{figVisualisation}
\end{figure}

\begin{figure}[h]
\includegraphics[width=90mm]{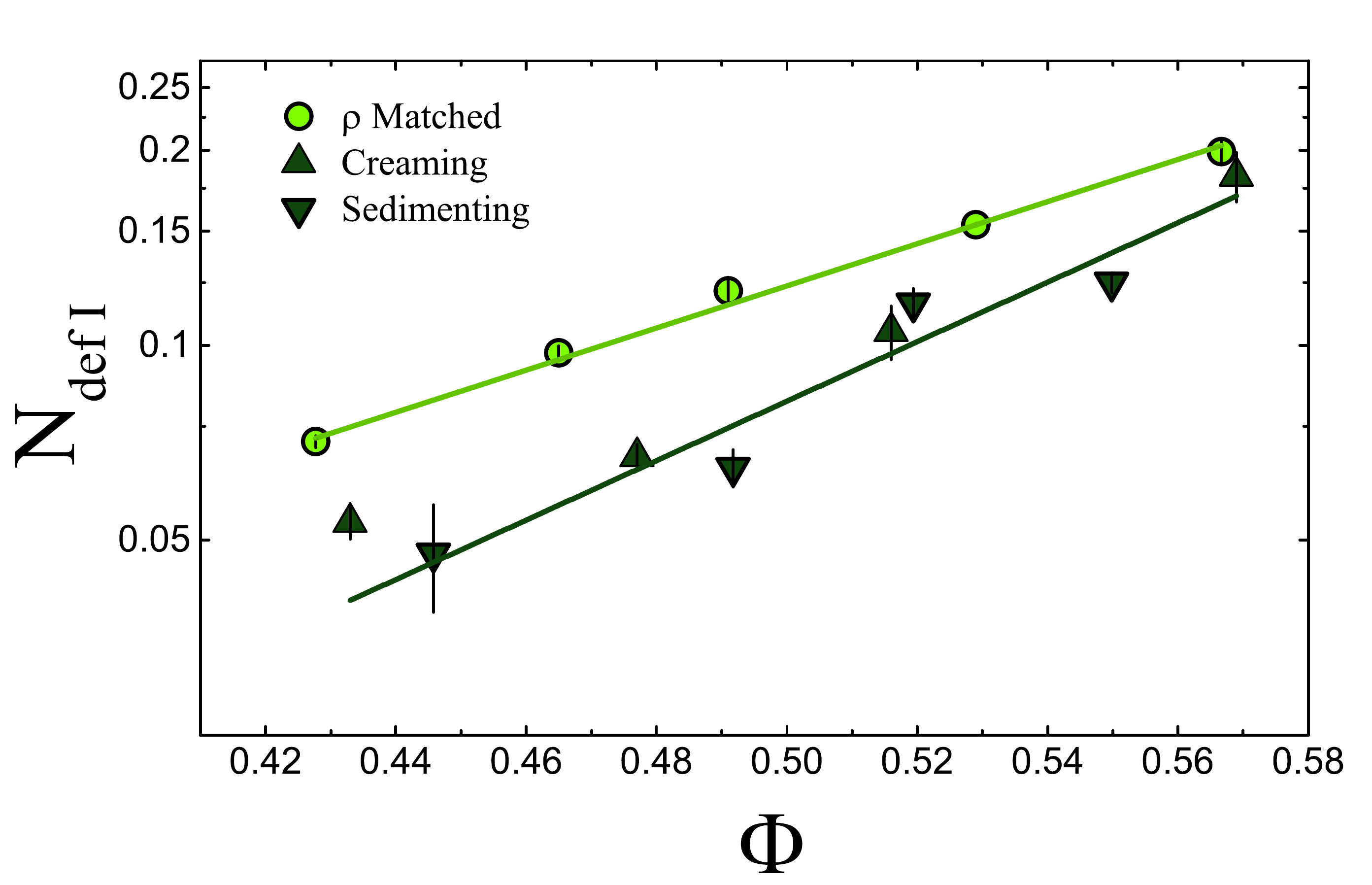}
\caption{\textbf{The effect of sedimentation on the structure of hard sphere fluids.} In the case of both sedimentation with a Pecl\'{e}t number of 1.5 and creaming with a Pecl\'{e}t number of 1.5 the population of our five fold symmetric structures is reduced by roughly a factor of 2.}
\label{fig10Bpopulation}
\end{figure}

Example experimental results are shown in Fig.~\ref{figVisualisation}, for a volume fraction of $\phi=0.45$. Here we see that, for a sedimentation rate of Pecl\'{e}t number Pe $\approx$ 1.5, the number of defective icosahedra is much reduced compared to the density matched case where sedimentation is negligible. We now show this behaviour across a more complete range of volume fractions, in Fig.~\ref{fig10Bpopulation}. It is important as well to confirm that the same behaviour is found in the case that the solvent mixture is both heavier and lighter than the colloids. This is because of the electrostatic interactions and imaging that might influence the delicate process of structural fluctuations depend on the solvent composition, as discussed in section \ref{expmethods}. That we see a drop in the population of defective icosahedra upon \emph{both} increasing and decreasing the colloid buoyancy gives us confidence that the effects we find are due to sedimentation and not to a change in the interactions between the particles. This analysis has been performed to match the simulation defective icosahedra populations, in line with previous work\cite{pinchaipat2017,hallett2018}.

We now consider the effect of a reduction of fivefold symmetry upon the nucleation barrier height. Recall that the experiments pertain to colloids sedimenting through a liquid medium which imparts hydrodynamic interactions between the particles. On the other hand, our rare event sampling uses Monte-Carlo simulation in which the effects of hydrodynamic interactions for a non-equilibrium phenomenon such as sedimentation have not been considered yet. While it has been possible to include hydrodynamic interactions in crystallising hard spheres at higher volume fraction (in the regime where there is no discrepancy) \cite{radu2014}, we believe coupling rare event sampling to non-equilibrium sedimentation would be an extremely challenging undertaking.

\begin{figure}
\includegraphics[width=80mm]{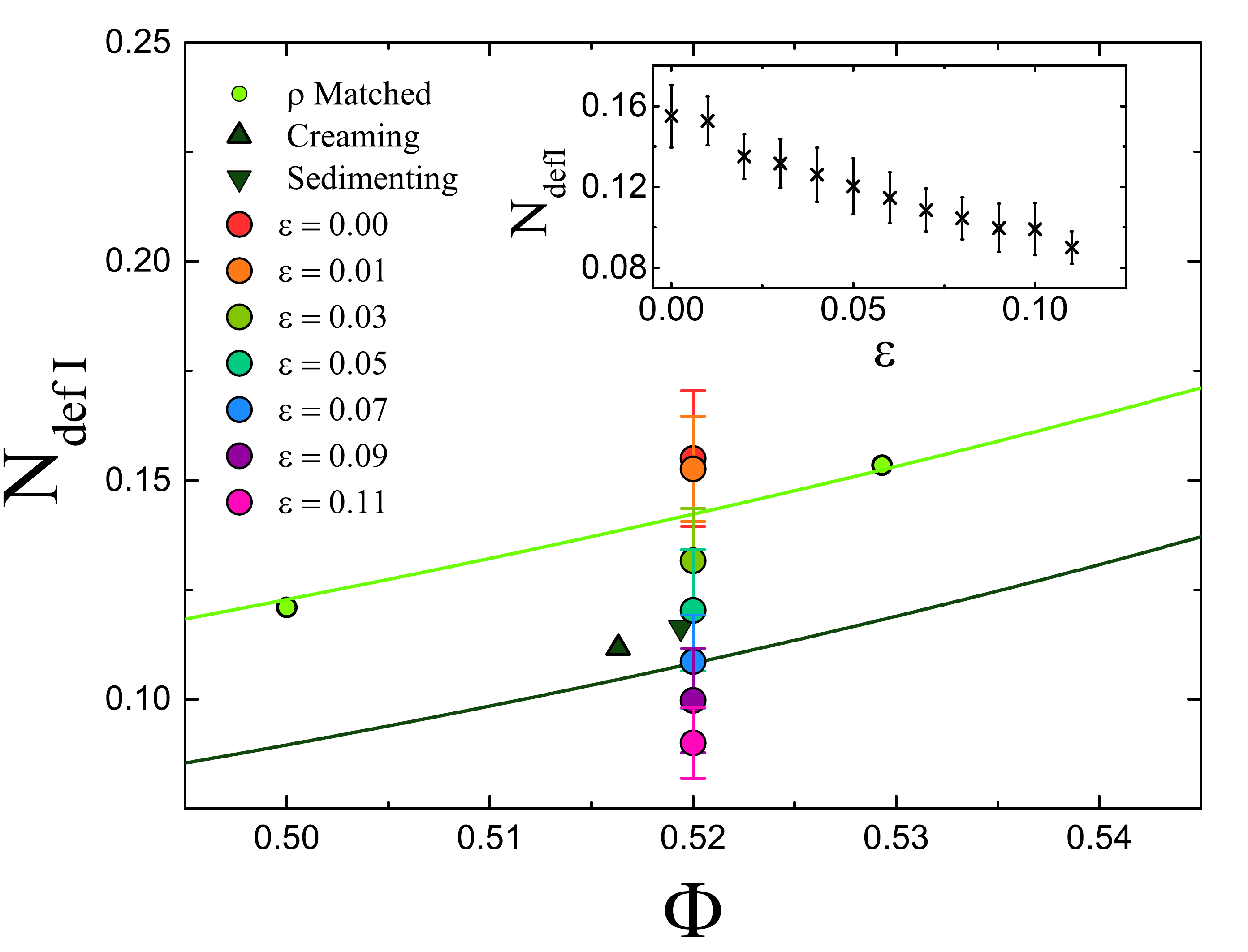}
\caption{\textbf{Effect of biasing against five fold symmetry.} In simulations, large field strengths tend to reduce the numbers of defective icosahedra. The field strength $\epsilon$ is tuned to bring about a reduction in the defective icosahedra equal to the reduction induced by sedimentation. We choose $\epsilon = 0.07$. The dark and light green lines are a zoom of Fig \ref{fig10Bpopulation} at the volume fractions which correspond to the supercooled region where the nucleation discrepancy exists}
\label{figMatchingFields}
\end{figure}

\begin{figure}
\includegraphics[width=80mm]{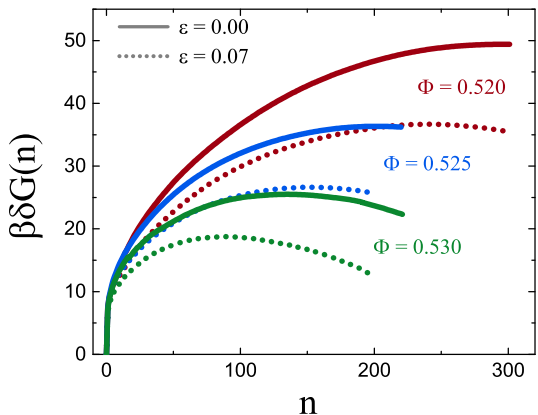}
\caption{\textbf{The effect of sedimentation on the nucleation barrier for hard spheres across three volume fractions.} The nucleation barrier height $\beta \delta G$ is plotted with respect to the size of the crystal nucleus, $n$. Filled circles show equilibrium simulations and empty circles show simulations with reduced five fold symmetry. The nucleation barriers in the case of reduced five fold symmetry are substantially lower than the unperturbed barriers, suggesting that sedimenting configurations are likely to nucleate much more quickly than unsedimenting configurations.}
\label{figBarriers}
\end{figure}

We therefore assume that it is the \emph{effect} of the hydrodynamic interactions that we observe, namely the reduction in fivefold symmetry, that is important, and investigate that with Monte-Carlo based umbrella sampling. To this end, we implement a model similar to that in Ref. 27. In particular, for every pentagonal bipyramid a particle is found in, we apply an energetic penalty, $\varepsilon$ $k_BT$. Pentagonal bipyramids are a smaller unit of fivefold symmetry, from which the defective icosahedra is built up\cite{taffs2016}. In order to achieve a comparable degree of suppression of defective icosahedra to that in the experiments, we gradually turn up the energetic penalty of forming one of these clusters and measure the effect on the number of defective icosahedra, as shown in Fig.~\ref{figMatchingFields} The value of the field that reproduces the experimentally observed change in the local structure populations is $\epsilon = 0.07$.

In Fig.~\ref{figBarriers} we show the barrier height obtained from umbrella sampling for the system in which the fivefold symmetry is suppressed. We find a that the nucleation barrier is suppressed by $11 k_BT$ at a volume fraction of $\Phi$=0.52, which increases the nucleation rates by $6.0\times 10^5$, all other factors being equal.
However, significant though this effect is, it does not resolve the entire discrepancy, because the Pecl\'{e}t number in our experiments is rather higher than those in the literature, as shown in table \ref{tableLiterature}.

\begin{table*}
\centering
\caption{Experimental systems previously investigated} 
\label{tableLiterature}
\begin{tabular}{|c|c|c|c|c|c|}
\hline
Author                    & Citation & Minimum phi & Pecl\'{e}t Number & Branch &Discrepancy size (Order of Magnitude)\\ \hline
Schatzel (1993) &  \cite{schatzel1993}        & 0.52        & 0.1868        & Strong Sedimentation & 12\\ \hline
He (1996)                 &   \cite{he1996}        & 0.522       & 0.0075        & Weak Sedimentation & 6  \\ \hline
He (1996)                 &    \cite{he1996}       & 0.517       & 0.1723        & Strong Sedimentation& 12 \\ \hline
Sinn (2001)               &   \cite{sinn2001}        & 0.518       & 0.179         & Strong Sedimentation & 10 \\ \hline
Harland (1997)            & \cite{harland1997}    & 0.53        & 0.006         & Weak Sedimentation  &3 ( $\phi = 0.53$) \\ \hline
Iacopini (2009)           &  \cite{iacopini2009}         & 0.535       & 0.018         & Weak Sedimentation & n/a  \\ \hline
Franke (2014)             & \cite{frankethesis}         & 0.525       & 0.009          & Weak Sedimentation &3 ($\phi = 0.53$)  \\ \hline
Taffs (2013)	& \cite{taffs2013}	& 0.54	& 0.001 \cite{royall2005s} & Very Weak Sedimentation & n/a \\ \hline
\end{tabular}
\end{table*}
\twocolumngrid

\begin{figure}
\includegraphics[width=80mm]{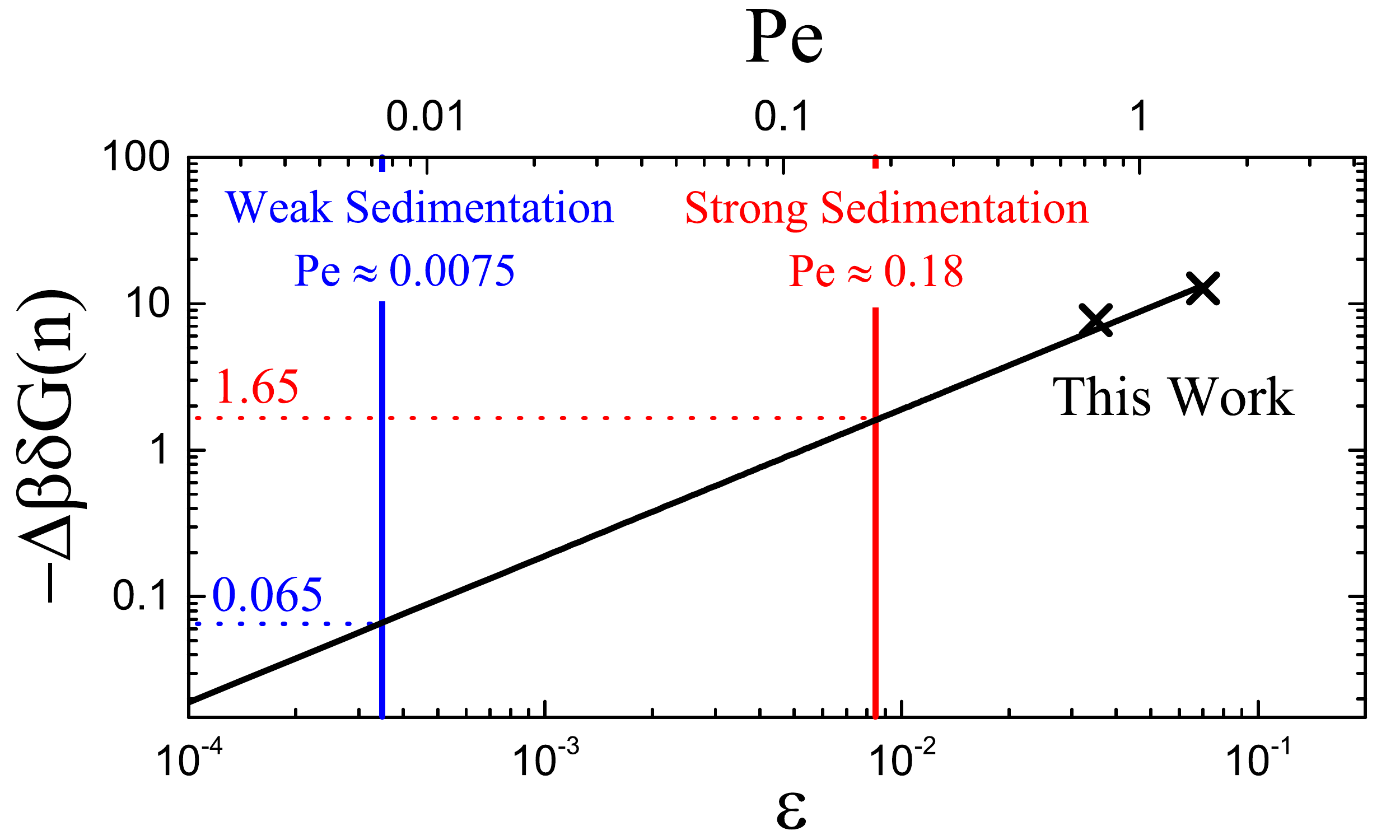}
\caption{\textbf{The change in barrier heights with respect to density matching} expected across a range of Pecl\'{e}t numbers. A linear fit is made to three points, the density matched case ($\pecl 
 = 0$), which of course sees no change in the barrier height, and barrier height changes at a field strength of $\epsilon = 0.07$ (shown in Fig.~\ref{figBarriers}) and  $\epsilon = 0.035$. The literature arm (see Fig.~\ref{figLiteratureNucleationRate}) which generally has weaker sedimentation and better agreement with simulations is hardly changed at all - a change in barrier height of $0.065 k_BT$, whilst the more strongly sedimenting data that agrees with simulation data poorly sees a change in barrier height of $1.65 k_BT$. This interpolation assumes a linear dependence of the defective icosahedra population on the Pecl\'{e}t numbers, and a linear dependence of the barrier height changes on the defective icosahedra population.}
\label{figPecletScaling}
\end{figure}

Figure \ref{figPecletScaling} shows how much the barrier heights would be expected to change
for the experiments shown in Fig. ~\ref{figLiteratureNucleationRate}, under the assumption that the fivefold symmetry is linear in Pecl\'{e}t number. The two data points shown from this work are changes in barrier heights caused by field strengths of $0.035 k_BT$ and $0.070 k_BT$. This data corresponds to unusually high Pecl\'{e}t numbers of 0.75 and 1.5, respectively, and the changes in barrier height scale approximately linearly with the field strength. By linearly interpolating between our data and the density-matched ($P_e\rightarrow0$) case to consider the Pecl\'{e}t numbers in the literature, we can see that the expected change in the barrier heights are $1.65 k_BT$ in the strongly sedimenting arm (shown in red-yellow in Fig.~\ref{figLiteratureNucleationRate}) and $0.065 k_BT$ in the weakly sedimenting arm (shown in blue-green in Fig.~\ref{figLiteratureNucleationRate}).

\subsection{Density Fluctuations}

A second structural aspect of our colloidal fluids that might also be altered by sedimentation is the distribution of the local densities --- the density fluctuations in the fluid. This has been suggested as a mechanism by which the nucleation rate might be enhanced in experiments relative to computer simulation\cite{russo2013} . A fluid with a broader distribution in local densities is expected to nucleate faster than one with a narrow distribution, as it will have more regions of high volume fraction in which critical nuclei will be preferentially born. To measure this quantity, for each particle we count the number of particles within a radius of \textit{r}= 3.25$\sigma$, calculate the resultant local volume fraction, and bin the local volume fractions of every particle. We chose this value of $r$ as it gives a mean value of the local densities equal to the bulk number density. This is because the radius of our sphere is a point where the $g(r)$ is equal to unity. 

\begin{figure*}
\includegraphics[width=175mm]{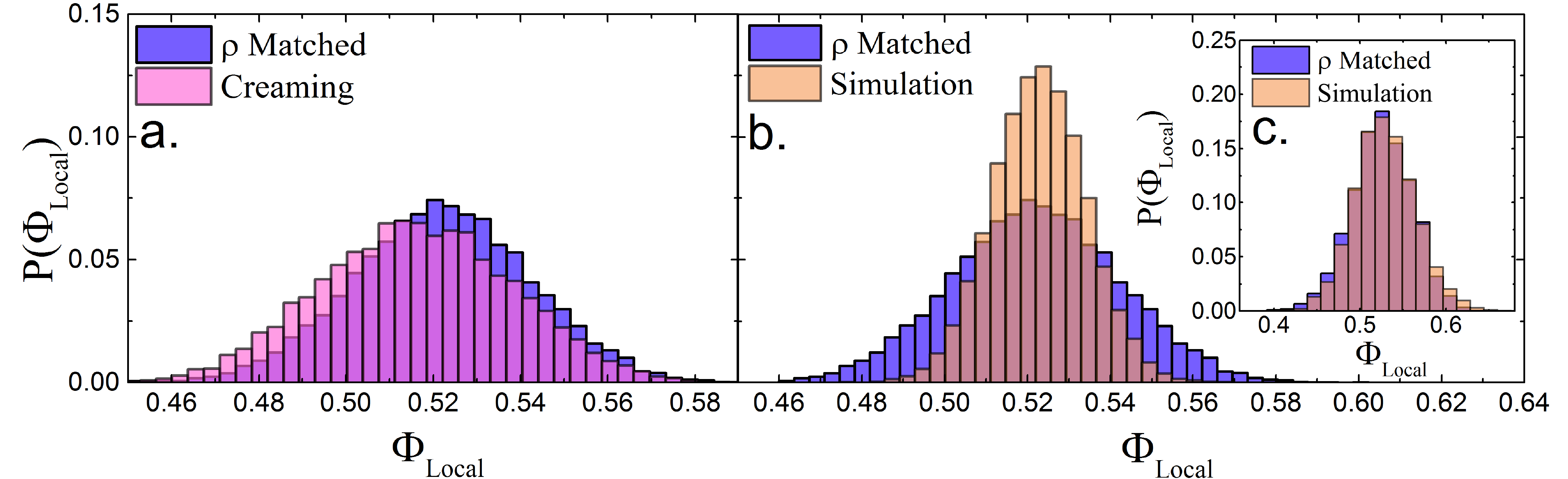}
\caption{\textbf{Comparing local density distributions.} \textbf{a.} The effect of sedimentation on density fluctuations is marginal --- here a CHB-rich case is shown with respect to the density matched case --- cis-decalin rich samples show similar behaviour. \textbf{b.} A much broader distribution of local number densities exists in 
experiments than in simulation (main figure). \textbf{c.} When the analysis is restrained to 2 dimensions this difference vanishes.}
\label{figsimvsexp}
\end{figure*}

This analysis shows that the standard deviation of the local volume fractions of our experimental system is around 0.02, whilst the standard deviation in our simulations is around 0.01. This difference gives a large increase in the number of particles which have high local densities in experiments, and as such might be a mechanism to partially explain the nucleation rate gap. Sedimentation makes essentially no difference to the density fluctuations. (see Fig.~\ref{figsimvsexp}).

However, this change in the density fluctuations can be understood as the consequence systematic experimental error when tracking 3d confocal images of colloids at high packing fraction rather than a true feature of our experimental hard sphere colloids. Initially we checked to see if these experimental fluctuations could be caused by slight changes in the interaction potential between the colloids, simulation ensemble or size, imaging parameters, polydispersity, or tracking parameters but none of these proved explanatory. We implement these possibilities in Monte Carlo simulations and the displayed results used the Colloids particle tracking package \cite{leocmach2013sm}, but other tracking codes, based ultimately on the methods introduced by Crocker and Grier \cite{crocker1995} recreated the fluctuation results. However, when we examine the fluctuations in two dimensions, in the $xy$ plane, they agree with simulation fluctuations in 2d. In confocal imaging, the $z$ dimension has a broader point spread function than $xy$, and this leads to artificially large fluctuations. When experiments and simulations at lower packing fraction are compared, good agreement can be found below $\phi = 0.40$, beyond this, the apparent fluctuations in the experiments diverge from the simulation result. We therefore conclude that examining fluctuation distributions in 3d constitutes a limitation in confocal colloidal experiments, at least at high packing fraction.

 \subsection{Decoupling Fluctuations and Local Order}

Given the aforementioned limitations in the particle tracking, we now address whether these systematic effects influence the dependence of the defective icosahedra populations on the Pecl\'et number (see Fig.~\ref{fig10Bpopulation}). We present three arguments. Firstly, as shown in Fig.~\ref{figsimvsexp}(a), the size of the density fluctuations does not depend on sedimentation, in contrast to the populations of defective icosahedra. Secondly, performing Monte Carlo simulations seeded with density matched experimental configurations, we observe that the initial experimentally measured values of the defective icosahedra population and the density fluctuations are at about $3\sigma$ and $30\sigma$ from the stationary values, respectively. 

\begin{figure}
\includegraphics[width=80mm]{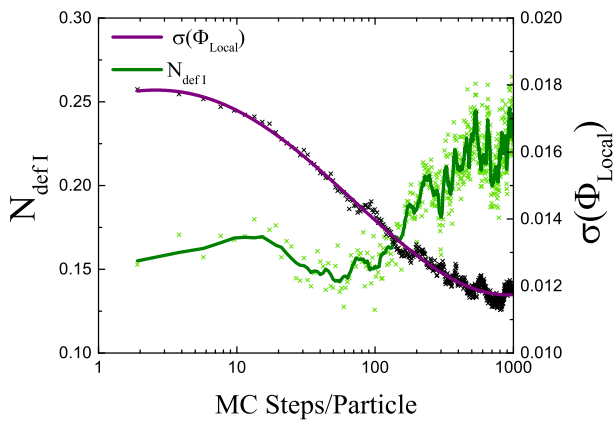}
\caption{\textbf{Monte Carlo relaxation of experimental configurations} Relaxing experimental configurations, we find that defective icosahedra populations remain fairly unaffected for small numbers of Monte Carlo steps (less than 200) whilst density fluctuations make a very rapid decline. This suggests that defective icosahedra populations are robust to the errors which cause the density fluctuation enlargement in experiment.}
\label{10B_Flucs_MCSim}
\end{figure}

Finally, to further demonstrate the robustness of the TCC analysis, in Fig.~\ref{10B_Flucs_MCSim} we plot both the evolution of density fluctuations and the population of defective icosahedra as a function of the number of steps for MC simulations that are started from experimentally extracted coordinates. The potential is chosen to be a soft sphere potential $V(r) = (\frac{\sigma}{r})^n$, with $n=20$. The hard sphere potential itself is not used as our initial experimental configurations contain some slightly overlapping particles due to particle tracking errors. To avoid boundary effects, we only consider the inner 1000 particles. We find that defective icosahedra populations remain fairly stable for small numbers of MC steps (fewer than 200) whilst density fluctuations make a very rapid decline. This suggests that defective icosahedra populations are robust to the errors which cause the increase in density fluctuations in experiment. Moreover, recently Bonn \textit{et al.}\cite{bonn2018} have also pointed out that nucleation rates seem to be unaffected by density fluctuations.

\section{Conclusions}

We have performed experiments examining the effect of sedimentation on the local structure of hard sphere colloidal fluids. This was motivated by the need to explain a huge discrepancy between the simulated and experimental nucleation rates of hard spheres at weak supercooling. We show that the degree of fivefold symmetry in hard sphere fluids is suppressed by sedimentation. This suppression of fivefold symmetry lowers the nucleation barrier and resolves a portion of the gap in nucleation rates. Further work should concentrate on the exact dependency between the degree of sedimentation and five fold symmetry. If this is non-linear, for example if the effect of sedimentation climbs rapidly at the low Pecl\'{e}t numbers relevant to earlier experiments, and then slows before a Pecl\'{e}t number $\pecl=1.5$ used here, this may explain a much larger portion of the discrepancy.

We also examine the effect of sedimentation on the local density fluctuations of the fluid. We find only a small difference between sedimenting and density matched systems. However, a very large discrepancy is found between all experiments and simulation. This is due to small systematic tracking errors created by relatively poorer resolution in the axial direction
and should serve as a cautionary tale of the limits of particle resolved colloidal experiments.

We conclude that while sedimentation will act to accelerate nucleation in hard sphere colloids, due to its effect on the fivefold symmetry in the metastable fluid, this seems unlikely to account for the extent of the discrepancy between experiment and simulation in this phenomenon. Our results suggest that density fluctuations in the sedimenting suspension should be unimportant. Given that effects such as polydispersity, electrostatic charge and hydrodynamic interactions \cite{auer2001,auer2002,radu2014} have been investigated, all to no avail, we conclude that the ``second-largest discrepancy in physics'' \cite{russo2013} remains unresolved.

\textbf{Acknowledgements ---- }
It is a pleasure to thank Daniele Coslovich, Daan Frenkel, Eduardo Sanz, Tanja Schilling, Thomas Speck, Hajime Tanaka and Chantal Valeriani for stimulating discussions. CPR acknowledges the Royal Society and European Research Council (ERC consolidator grant NANOPRS, project number 617266) for financial support. NW is supported by EPSRC grant code EP/E025377/1. J.R. acknowledges support from the European Research Council Grant DLV-759187 and the Royal Society University Research Fellowship. This work was carried out using the facilities of the Advanced Computing Research Centre, University of Bristol.


\begin{thebibliography}{42}
\expandafter\ifx\csname natexlab\endcsname\relax\def\natexlab#1{#1}\fi
\expandafter\ifx\csname bibnamefont\endcsname\relax
  \def\bibnamefont#1{#1}\fi
\expandafter\ifx\csname bibfnamefont\endcsname\relax
  \def\bibfnamefont#1{#1}\fi
\expandafter\ifx\csname citenamefont\endcsname\relax
  \def\citenamefont#1{#1}\fi
\expandafter\ifx\csname url\endcsname\relax
  \def\url#1{\texttt{#1}}\fi
\expandafter\ifx\csname urlprefix\endcsname\relax\def\urlprefix{URL }\fi
\providecommand{\bibinfo}[2]{#2}
\providecommand{\eprint}[2][]{\url{#2}}

\bibitem[{\citenamefont{Sear}(2007)}]{sear2007}
\bibinfo{author}{\bibfnamefont{R.~P.} \bibnamefont{Sear}}, \bibinfo{journal}{J.
  Phys.: Condens. Matter} \textbf{\bibinfo{volume}{19}},
  \bibinfo{pages}{033101} (\bibinfo{year}{2007}).

\bibitem[{\citenamefont{Schilling et~al.}(2011)\citenamefont{Schilling, Dorosz,
  Schoepe, and Opletal}}]{schilling2011}
\bibinfo{author}{\bibfnamefont{T.}~\bibnamefont{Schilling}},
  \bibinfo{author}{\bibfnamefont{S.}~\bibnamefont{Dorosz}},
  \bibinfo{author}{\bibfnamefont{H.~J.} \bibnamefont{Schoepe}},
  \bibnamefont{and} \bibinfo{author}{\bibfnamefont{G.}~\bibnamefont{Opletal}},
  \bibinfo{journal}{J. Phys.: Condens. Matter} \textbf{\bibinfo{volume}{23}},
  \bibinfo{pages}{194120} (\bibinfo{year}{2011}).

\bibitem[{\citenamefont{Hunter and Weeks}(2012)}]{hunter2012}
\bibinfo{author}{\bibfnamefont{G.~L.} \bibnamefont{Hunter}} \bibnamefont{and}
  \bibinfo{author}{\bibfnamefont{E.~R.} \bibnamefont{Weeks}},
  \bibinfo{journal}{Rep. Prog. Phys.} \textbf{\bibinfo{volume}{75}},
  \bibinfo{pages}{066501} (\bibinfo{year}{2012}).

\bibitem[{\citenamefont{Royall et~al.}(2013)\citenamefont{Royall, Poon, and
  Weeks}}]{royall2013myth}
\bibinfo{author}{\bibfnamefont{C.~P.} \bibnamefont{Royall}},
  \bibinfo{author}{\bibfnamefont{W.~C.~K.} \bibnamefont{Poon}},
  \bibnamefont{and} \bibinfo{author}{\bibfnamefont{E.~R.} \bibnamefont{Weeks}},
  \bibinfo{journal}{Soft Matter} \textbf{\bibinfo{volume}{9}},
  \bibinfo{pages}{17} (\bibinfo{year}{2013}).

\bibitem[{\citenamefont{Palberg}(2014)}]{palberg2014}
\bibinfo{author}{\bibfnamefont{T.}~\bibnamefont{Palberg}}, \bibinfo{journal}{J
  Phys.: Condens Matter} \textbf{\bibinfo{volume}{26}}, \bibinfo{pages}{333101}
  (\bibinfo{year}{2014}).

\bibitem[{\citenamefont{Gasser et~al.}(2001)\citenamefont{Gasser, Weeks,
  Schofield, Pusey, and Weitz}}]{gasser2001}
\bibinfo{author}{\bibfnamefont{U.}~\bibnamefont{Gasser}},
  \bibinfo{author}{\bibfnamefont{E.~R.} \bibnamefont{Weeks}},
  \bibinfo{author}{\bibfnamefont{A.}~\bibnamefont{Schofield}},
  \bibinfo{author}{\bibfnamefont{P.~N.} \bibnamefont{Pusey}}, \bibnamefont{and}
  \bibinfo{author}{\bibfnamefont{D.~A.} \bibnamefont{Weitz}},
  \bibinfo{journal}{Science} \textbf{\bibinfo{volume}{292}},
  \bibinfo{pages}{258} (\bibinfo{year}{2001}).

\bibitem[{\citenamefont{Gasser}(2009)}]{gasser2009}
\bibinfo{author}{\bibfnamefont{U.}~\bibnamefont{Gasser}}, \bibinfo{journal}{J.
  Phys.: Condens. Matter} \textbf{\bibinfo{volume}{21}},
  \bibinfo{pages}{203101} (\bibinfo{year}{2009}).

\bibitem[{\citenamefont{Ivlev et~al.}(2012)\citenamefont{Ivlev, L\"{o}wen,
  Morfill, and Royall}}]{ivlev}
\bibinfo{author}{\bibfnamefont{A.}~\bibnamefont{Ivlev}},
  \bibinfo{author}{\bibfnamefont{H.}~\bibnamefont{L\"{o}wen}},
  \bibinfo{author}{\bibfnamefont{G.~E.} \bibnamefont{Morfill}},
  \bibnamefont{and} \bibinfo{author}{\bibfnamefont{C.~P.}
  \bibnamefont{Royall}}, \emph{\bibinfo{title}{Complex Plasmas and Colloidal
  Dispersions: Particle-resolved Studies of Classical Liquids and Solids}}
  (\bibinfo{publisher}{World Scientific Publishing Co., Singapore Scientific},
  \bibinfo{year}{2012}).

\bibitem[{\citenamefont{Allen et~al.}(2009)\citenamefont{Allen, Valeriani, and
  ten Wolde}}]{allen2009}
\bibinfo{author}{\bibfnamefont{R.~J.} \bibnamefont{Allen}},
  \bibinfo{author}{\bibfnamefont{C.}~\bibnamefont{Valeriani}},
  \bibnamefont{and} \bibinfo{author}{\bibfnamefont{P.~R.} \bibnamefont{ten
  Wolde}}, \bibinfo{journal}{J. Phys. Cond. Matt.}
  \textbf{\bibinfo{volume}{21}}, \bibinfo{pages}{463102}
  (\bibinfo{year}{2009}).

\bibitem[{\citenamefont{Ghaemi et~al.}(2012)\citenamefont{Ghaemi, Minozzi,
  Carloni, and Laio}}]{ghaemi2012}
\bibinfo{author}{\bibfnamefont{Z.}~\bibnamefont{Ghaemi}},
  \bibinfo{author}{\bibfnamefont{M.}~\bibnamefont{Minozzi}},
  \bibinfo{author}{\bibfnamefont{P.}~\bibnamefont{Carloni}}, \bibnamefont{and}
  \bibinfo{author}{\bibfnamefont{A.}~\bibnamefont{Laio}}, \bibinfo{journal}{The
  Journal of Physical Chemistry B} \textbf{\bibinfo{volume}{116}},
  \bibinfo{pages}{8714} (\bibinfo{year}{2012}).

\bibitem[{\citenamefont{Bussi et~al.}(2006)\citenamefont{Bussi, Gervasio, Laio,
  and Parrinello}}]{bussi2006}
\bibinfo{author}{\bibfnamefont{G.}~\bibnamefont{Bussi}},
  \bibinfo{author}{\bibfnamefont{F.~L.} \bibnamefont{Gervasio}},
  \bibinfo{author}{\bibfnamefont{A.}~\bibnamefont{Laio}}, \bibnamefont{and}
  \bibinfo{author}{\bibfnamefont{M.}~\bibnamefont{Parrinello}},
  \bibinfo{journal}{Journal of the American Chemical Society}
  \textbf{\bibinfo{volume}{128}}, \bibinfo{pages}{13435}
  (\bibinfo{year}{2006}).

\bibitem[{\citenamefont{Auer and Frenkel}(2001)}]{auer2001}
\bibinfo{author}{\bibfnamefont{S.}~\bibnamefont{Auer}} \bibnamefont{and}
  \bibinfo{author}{\bibfnamefont{D.}~\bibnamefont{Frenkel}},
  \bibinfo{journal}{Nature} \textbf{\bibinfo{volume}{409}},
  \bibinfo{pages}{1020} (\bibinfo{year}{2001}).

\bibitem[{\citenamefont{Auer and Frenkel}(2004)}]{auer2004}
\bibinfo{author}{\bibfnamefont{S.}~\bibnamefont{Auer}} \bibnamefont{and}
  \bibinfo{author}{\bibfnamefont{D.}~\bibnamefont{Frenkel}},
  \bibinfo{journal}{Annu. Rev. Phys. Chem.} \textbf{\bibinfo{volume}{55}},
  \bibinfo{pages}{333} (\bibinfo{year}{2004}).

\bibitem[{\citenamefont{Filion et~al.}(2011)\citenamefont{Filion, Ni, Frenkel,
  and Dijkstra}}]{filion2011}
\bibinfo{author}{\bibfnamefont{L.}~\bibnamefont{Filion}},
  \bibinfo{author}{\bibfnamefont{R.}~\bibnamefont{Ni}},
  \bibinfo{author}{\bibfnamefont{D.}~\bibnamefont{Frenkel}}, \bibnamefont{and}
  \bibinfo{author}{\bibfnamefont{M.}~\bibnamefont{Dijkstra}},
  \bibinfo{journal}{J. Chem. Phys} \textbf{\bibinfo{volume}{134}},
  \bibinfo{pages}{134901} (\bibinfo{year}{2011}).

\bibitem[{\citenamefont{Auer and Frenkel}(2002)}]{auer2002}
\bibinfo{author}{\bibfnamefont{S.}~\bibnamefont{Auer}} \bibnamefont{and}
  \bibinfo{author}{\bibfnamefont{D.}~\bibnamefont{Frenkel}},
  \bibinfo{journal}{J. Phys.: Condens. Matter} \textbf{\bibinfo{volume}{14}},
  \bibinfo{pages}{7667} (\bibinfo{year}{2002}).

\bibitem[{\citenamefont{Radu and Schilling}(2014)}]{radu2014}
\bibinfo{author}{\bibfnamefont{M.}~\bibnamefont{Radu}} \bibnamefont{and}
  \bibinfo{author}{\bibfnamefont{T.}~\bibnamefont{Schilling}},
  \bibinfo{journal}{EPL (Europhysics Letters)} \textbf{\bibinfo{volume}{105}},
  \bibinfo{pages}{26001} (\bibinfo{year}{2014}).

\bibitem[{\citenamefont{Russo et~al.}(2013)\citenamefont{Russo, Maggs, Bonn,
  and Tanaka}}]{russo2013}
\bibinfo{author}{\bibfnamefont{J.}~\bibnamefont{Russo}},
  \bibinfo{author}{\bibfnamefont{A.~C.} \bibnamefont{Maggs}},
  \bibinfo{author}{\bibfnamefont{D.}~\bibnamefont{Bonn}}, \bibnamefont{and}
  \bibinfo{author}{\bibfnamefont{H.}~\bibnamefont{Tanaka}},
  \bibinfo{journal}{Soft Matter} \textbf{\bibinfo{volume}{9}},
  \bibinfo{pages}{7369} (\bibinfo{year}{2013}).

\bibitem[{\citenamefont{Sear}(2012)}]{sear2012}
\bibinfo{author}{\bibfnamefont{R.~P.} \bibnamefont{Sear}},
  \bibinfo{journal}{Int. Mater. Rev.} \textbf{\bibinfo{volume}{57}},
  \bibinfo{pages}{328} (\bibinfo{year}{2012}).

\bibitem[{\citenamefont{Taffs et~al.}(2013)\citenamefont{Taffs, Williams,
  Tanaka, and Royall}}]{taffs2013}
\bibinfo{author}{\bibfnamefont{J.}~\bibnamefont{Taffs}},
  \bibinfo{author}{\bibfnamefont{S.~R.} \bibnamefont{Williams}},
  \bibinfo{author}{\bibfnamefont{H.}~\bibnamefont{Tanaka}}, \bibnamefont{and}
  \bibinfo{author}{\bibfnamefont{C.~P.} \bibnamefont{Royall}},
  \bibinfo{journal}{Soft Matter} \textbf{\bibinfo{volume}{9}},
  \bibinfo{pages}{297} (\bibinfo{year}{2013}).

\bibitem[{\citenamefont{Schätzel and Ackerson}(1993)}]{schatzel1993}
\bibinfo{author}{\bibfnamefont{K.}~\bibnamefont{Schätzel}} \bibnamefont{and}
  \bibinfo{author}{\bibfnamefont{B.~J.} \bibnamefont{Ackerson}},
  \bibinfo{journal}{Physical Review E} \textbf{\bibinfo{volume}{48}},
  \bibinfo{pages}{3766, 3777} (\bibinfo{year}{1993}).

\bibitem[{\citenamefont{He et~al.}(1996)\citenamefont{He, Ackerson, van Megen,
  Underwood, and Sch\"atzel}}]{he1996}
\bibinfo{author}{\bibfnamefont{Y.}~\bibnamefont{He}},
  \bibinfo{author}{\bibfnamefont{B.~J.} \bibnamefont{Ackerson}},
  \bibinfo{author}{\bibfnamefont{W.}~\bibnamefont{van Megen}},
  \bibinfo{author}{\bibfnamefont{S.~M.} \bibnamefont{Underwood}},
  \bibnamefont{and}
  \bibinfo{author}{\bibfnamefont{K.}~\bibnamefont{Sch\"atzel}},
  \bibinfo{journal}{Phys. Rev. E} \textbf{\bibinfo{volume}{54}},
  \bibinfo{pages}{5286} (\bibinfo{year}{1996}).

\bibitem[{\citenamefont{Sinn et~al.}(2001)\citenamefont{Sinn, Heymann, Stipp,
  and Palberg}}]{sinn2001}
\bibinfo{author}{\bibfnamefont{C.}~\bibnamefont{Sinn}},
  \bibinfo{author}{\bibfnamefont{A.}~\bibnamefont{Heymann}},
  \bibinfo{author}{\bibfnamefont{A.}~\bibnamefont{Stipp}}, \bibnamefont{and}
  \bibinfo{author}{\bibfnamefont{T.}~\bibnamefont{Palberg}},
  \bibinfo{journal}{Progress in Colloid and Polymer Science}
  \textbf{\bibinfo{volume}{118}}, \bibinfo{pages}{266} (\bibinfo{year}{2001}).

\bibitem[{\citenamefont{Harland and van Megen}(1997)}]{harland1997}
\bibinfo{author}{\bibfnamefont{J.~L.} \bibnamefont{Harland}} \bibnamefont{and}
  \bibinfo{author}{\bibfnamefont{W.}~\bibnamefont{van Megen}},
  \bibinfo{journal}{Phys. Rev. E.} \textbf{\bibinfo{volume}{55}},
  \bibinfo{pages}{3054} (\bibinfo{year}{1997}).

\bibitem[{\citenamefont{Iacopini et~al.}(2009)\citenamefont{Iacopini, Palberg,
  and Sch\"ope}}]{iacopini2009}
\bibinfo{author}{\bibfnamefont{S.}~\bibnamefont{Iacopini}},
  \bibinfo{author}{\bibfnamefont{T.}~\bibnamefont{Palberg}}, \bibnamefont{and}
  \bibinfo{author}{\bibfnamefont{H.~J.} \bibnamefont{Sch\"ope}},
  \bibinfo{journal}{Phys. Rev. E} \textbf{\bibinfo{volume}{79}},
  \bibinfo{pages}{010601} (\bibinfo{year}{2009}).

\bibitem[{\citenamefont{Franke}(2014)}]{frankethesis}
\bibinfo{author}{\bibfnamefont{M.}~\bibnamefont{Franke}},
  \bibinfo{journal}{Thesis, Mainz}  (\bibinfo{year}{2014}).

\bibitem[{\citenamefont{Frank}(1952)}]{frank1952}
\bibinfo{author}{\bibfnamefont{F.~C.} \bibnamefont{Frank}},
  \bibinfo{journal}{Proc. R. Soc. A.} \textbf{\bibinfo{volume}{215}},
  \bibinfo{pages}{43} (\bibinfo{year}{1952}).

\bibitem[{\citenamefont{Taffs and Patrick~Royall}(2016)}]{taffs2016}
\bibinfo{author}{\bibfnamefont{J.}~\bibnamefont{Taffs}} \bibnamefont{and}
  \bibinfo{author}{\bibfnamefont{C.}~\bibnamefont{Patrick~Royall}},
  \bibinfo{journal}{Nature Communications} \textbf{\bibinfo{volume}{7}},
  \bibinfo{pages}{13225 EP } (\bibinfo{year}{2016}).

\bibitem[{\citenamefont{Malins et~al.}(2013)\citenamefont{Malins, Williams,
  Eggers, and Royall}}]{malins2013tcc}
\bibinfo{author}{\bibfnamefont{A.}~\bibnamefont{Malins}},
  \bibinfo{author}{\bibfnamefont{S.~R.} \bibnamefont{Williams}},
  \bibinfo{author}{\bibfnamefont{J.}~\bibnamefont{Eggers}}, \bibnamefont{and}
  \bibinfo{author}{\bibfnamefont{C.~P.} \bibnamefont{Royall}},
  \bibinfo{journal}{J. Chem. Phys.} \textbf{\bibinfo{volume}{139}},
  \bibinfo{pages}{234506} (\bibinfo{year}{2013}).

\bibitem[{\citenamefont{Royall et~al.}(2015)\citenamefont{Royall, Eggers,
  Furukawa, and Tanaka}}]{royall2015}
\bibinfo{author}{\bibfnamefont{C.~P.} \bibnamefont{Royall}},
  \bibinfo{author}{\bibfnamefont{J.}~\bibnamefont{Eggers}},
  \bibinfo{author}{\bibfnamefont{A.}~\bibnamefont{Furukawa}}, \bibnamefont{and}
  \bibinfo{author}{\bibfnamefont{H.}~\bibnamefont{Tanaka}},
  \bibinfo{journal}{Phys. Rev. Lett.} \textbf{\bibinfo{volume}{114}},
  \bibinfo{pages}{258302} (\bibinfo{year}{2015}).

\bibitem[{\citenamefont{Royall and Kob}(2017)}]{royall2017}
\bibinfo{author}{\bibfnamefont{C.~P.} \bibnamefont{Royall}} \bibnamefont{and}
  \bibinfo{author}{\bibfnamefont{W.}~\bibnamefont{Kob}}, \bibinfo{journal}{J.
  Stat. Mech.: Theory and Experiment} p. \bibinfo{pages}{024001}
  (\bibinfo{year}{2017}).

\bibitem[{\citenamefont{Royall et~al.}(2018)\citenamefont{Royall, Williams, and
  Tanaka}}]{royall2018}
\bibinfo{author}{\bibfnamefont{C.~P.} \bibnamefont{Royall}},
  \bibinfo{author}{\bibfnamefont{S.~R.} \bibnamefont{Williams}},
  \bibnamefont{and} \bibinfo{author}{\bibfnamefont{H.}~\bibnamefont{Tanaka}},
  \bibinfo{journal}{J. Chem. Phys.} \textbf{\bibinfo{volume}{148}},
  \bibinfo{pages}{044501} (\bibinfo{year}{2018}).

\bibitem[{\citenamefont{Hansen and Macdonald}(1976)}]{hansen}
\bibinfo{author}{\bibfnamefont{J.-P.} \bibnamefont{Hansen}} \bibnamefont{and}
  \bibinfo{author}{\bibfnamefont{I.}~\bibnamefont{Macdonald}},
  \emph{\bibinfo{title}{Theory of Simple Liquids}}
  (\bibinfo{publisher}{Academic, London}, \bibinfo{year}{1976}).

\bibitem[{\citenamefont{Pinchaipat et~al.}(2017)\citenamefont{Pinchaipat,
  Campo, Turci, Hallett, Speck, and Royall}}]{pinchaipat2017}
\bibinfo{author}{\bibfnamefont{R.}~\bibnamefont{Pinchaipat}},
  \bibinfo{author}{\bibfnamefont{M.}~\bibnamefont{Campo}},
  \bibinfo{author}{\bibfnamefont{F.}~\bibnamefont{Turci}},
  \bibinfo{author}{\bibfnamefont{J.~E.} \bibnamefont{Hallett}},
  \bibinfo{author}{\bibfnamefont{T.}~\bibnamefont{Speck}}, \bibnamefont{and}
  \bibinfo{author}{\bibfnamefont{C.~P.} \bibnamefont{Royall}},
  \bibinfo{journal}{Phys. Rev. Lett.} \textbf{\bibinfo{volume}{119}},
  \bibinfo{pages}{028004} (\bibinfo{year}{2017}).

\bibitem[{\citenamefont{Hallet et~al.}(2018)\citenamefont{Hallet, Turci, and
  Royall}}]{hallett2018}
\bibinfo{author}{\bibfnamefont{J.~E.} \bibnamefont{Hallet}},
  \bibinfo{author}{\bibfnamefont{F.}~\bibnamefont{Turci}}, \bibnamefont{and}
  \bibinfo{author}{\bibfnamefont{C.~P.} \bibnamefont{Royall}},
  \bibinfo{journal}{accepted by Nature Comms.}  (\bibinfo{year}{2018}).

\bibitem[{\citenamefont{Sanz and Marenduzzo}(2010)}]{sanz2010}
\bibinfo{author}{\bibfnamefont{E.}~\bibnamefont{Sanz}} \bibnamefont{and}
  \bibinfo{author}{\bibfnamefont{D.}~\bibnamefont{Marenduzzo}},
  \bibinfo{journal}{J. Chem. Phys.} \textbf{\bibinfo{volume}{192}},
  \bibinfo{pages}{194102} (\bibinfo{year}{2010}).

\bibitem[{\citenamefont{Honeycutt and Andersen}(1987)}]{honeycutt1987}
\bibinfo{author}{\bibfnamefont{J.~D.} \bibnamefont{Honeycutt}}
  \bibnamefont{and} \bibinfo{author}{\bibfnamefont{H.~C.}
  \bibnamefont{Andersen}}, \bibinfo{journal}{J. Phys. Chem.}
  \textbf{\bibinfo{volume}{91}}, \bibinfo{pages}{4950} (\bibinfo{year}{1987}).

\bibitem[{\citenamefont{Rovigatti et~al.}(2018)\citenamefont{Rovigatti, Russo,
  and Romano}}]{rovigatti2018}
\bibinfo{author}{\bibfnamefont{L.}~\bibnamefont{Rovigatti}},
  \bibinfo{author}{\bibfnamefont{J.}~\bibnamefont{Russo}}, \bibnamefont{and}
  \bibinfo{author}{\bibfnamefont{F.}~\bibnamefont{Romano}},
  \bibinfo{journal}{Eur. Phys. J. E.} \textbf{\bibinfo{volume}{41}},
  \bibinfo{pages}{59} (\bibinfo{year}{2018}), ISSN \bibinfo{issn}{1292-895X}.

\bibitem[{\citenamefont{Poon et~al.}(2012)\citenamefont{Poon, Weeks, and
  Royall}}]{poon2012}
\bibinfo{author}{\bibfnamefont{W.~C.~K.} \bibnamefont{Poon}},
  \bibinfo{author}{\bibfnamefont{E.~R.} \bibnamefont{Weeks}}, \bibnamefont{and}
  \bibinfo{author}{\bibfnamefont{C.~P.} \bibnamefont{Royall}},
  \bibinfo{journal}{Soft Matter} \textbf{\bibinfo{volume}{8}},
  \bibinfo{pages}{21} (\bibinfo{year}{2012}).

\bibitem[{\citenamefont{Royall et~al.}(2005)\citenamefont{Royall, van Roij, and
  van Blaaderen}}]{royall2005s}
\bibinfo{author}{\bibfnamefont{C.}~\bibnamefont{Royall}},
  \bibinfo{author}{\bibfnamefont{R.}~\bibnamefont{van Roij}}, \bibnamefont{and}
  \bibinfo{author}{\bibfnamefont{A.}~\bibnamefont{van Blaaderen}},
  \bibinfo{journal}{J. Phys.: Condens. Matter} \textbf{\bibinfo{volume}{17}},
  \bibinfo{pages}{2315} (\bibinfo{year}{2005}).

\bibitem[{\citenamefont{Leocmach and Tanaka}(2013)}]{leocmach2013sm}
\bibinfo{author}{\bibfnamefont{M.}~\bibnamefont{Leocmach}} \bibnamefont{and}
  \bibinfo{author}{\bibfnamefont{H.}~\bibnamefont{Tanaka}},
  \bibinfo{journal}{Soft Matter} \textbf{\bibinfo{volume}{9}},
  \bibinfo{pages}{1447} (\bibinfo{year}{2013}).

\bibitem[{\citenamefont{Crocker and Grier}(1995)}]{crocker1995}
\bibinfo{author}{\bibfnamefont{J.~C.} \bibnamefont{Crocker}} \bibnamefont{and}
  \bibinfo{author}{\bibfnamefont{D.~G.} \bibnamefont{Grier}},
  \bibinfo{journal}{J. Coll. Interf. Sci.} \textbf{\bibinfo{volume}{179}},
  \bibinfo{pages}{298} (\bibinfo{year}{1995}).

\bibitem[{\citenamefont{Ketzetzi et~al.}(2018)\citenamefont{Ketzetzi, Russo,
  and Bonn}}]{bonn2018}
\bibinfo{author}{\bibfnamefont{S.}~\bibnamefont{Ketzetzi}},
  \bibinfo{author}{\bibfnamefont{J.}~\bibnamefont{Russo}}, \bibnamefont{and}
  \bibinfo{author}{\bibfnamefont{D.}~\bibnamefont{Bonn}}, \bibinfo{journal}{The
  Journal of Chemical Physics} \textbf{\bibinfo{volume}{148}},
  \bibinfo{pages}{064901} (\bibinfo{year}{2018}).

\end{thebibliography}

\end{document}